\documentclass[prb,twocolumn,showpacs,superscriptaddress,preprintnumbers,amsmath,amssymb,floatfix]{revtex4}

\usepackage{dcolumn}
\usepackage{graphicx}
\usepackage{bm} 

\newcommand{\eM}     {$\epsilon$-machine}
\newcommand{\eMs}    {$\epsilon$-machines}
\newcommand{\EM}     {$\epsilon$-Machine}
\newcommand{\EMs}    {$\epsilon$-Machines}

\newcommand{\CausalState}	{ {\cal S} }

\newcommand{\Prob}		{ {\rm P}}

\newcommand{\Cmu}		{ {C_\mu}}
\newcommand{\hmu}		{ {h_\mu}}
\newcommand{\EE}		{ {\bf E}}

\newcommand{\Range} {{r}}

\newcommand{\eMSR}    {{{$\epsilon$}MSR}}

\begin{document}

\title{Inferring Pattern and Disorder in\\
Close-Packed Structures from X-ray Diffraction Studies, Part II:\\
Structure and Intrinsic Computation in Zinc Sulphide}

\author{D. P. Varn}
\affiliation{Santa Fe Institute, 1399 Hyde Park Road, Santa Fe,
New Mexico 87501}
\affiliation{Department of Physics and Astronomy, University of Tennessee,
Knoxville, Tennessee 37996}

\author{G. S. Canright}
\affiliation{Department of Physics and Astronomy, University of Tennessee,
Knoxville, Tennessee 37996}
\affiliation{Telenor Research and Development, 1331 Fornebu, Norway}

\author{J. P. Crutchfield}
\affiliation{Santa Fe Institute, 1399 Hyde Park Road, Santa Fe,
New Mexico 87501}

\date{\today}

\begin{abstract}
In the previous paper of this series [D. P. Varn, G. S. Canright, and J.
P. Crutchfield, Physical Review B, \emph{submitted}] we detailed a
procedure---\emph{\eM\ spectral reconstruction}---to discover and
analyze patterns and disorder in close-packed structures as revealed
in x-ray diffraction spectra. We argued that this
\emph{computational mechanics} approach is more general than the
current alternative theory, the \emph{fault model}, and that it
provides a unique characterization of the disorder present. We
demonstrated the efficacy of computational mechanics on four prototype
spectra, finding that it was able to recover a statistical description
of the underlying modular-layer stacking using \eM\ representations.
Here we use this procedure to analyze structure and disorder in
four previously published zinc sulphide diffraction spectra. We
selected zinc sulphide not only for the theoretical interest this
material has attracted in an effort to develop an understanding of polytypism, but 
also because it displays solid-state phase transitions and 
experimental data is available. With the first spectrum we find 
qualitative agreement with earlier fault-model analyses, although
the reconstructed \eM\ detects structures not previously observed.
In the second spectrum, the results cannot be expressed in terms of
weak faulting and so no direct comparison between the fault model
and the reconstructed \eM\ is possible. Nonetheless, we show that
the \eM\ gives substantially better experimental agreement and a
number of structural insights. In the third spectrum, the fault
model fails completely due to the high degree of disorder present,
while the reconstructed \eM\ reproduces the experimental spectrum
well. In the fourth spectrum, we again find good quantitative
agreement with experiment but find that the \eM\ has difficulty reproducing
the shape of several Bragg-like peaks. We discuss the reasons for this. 
Using the \eMs\ reconstructed for each spectrum, we calculate
a number of physical parameters---such as, stacking energies,
configurational entropies, and hexagonality---and several
quantities---including statistical complexity and excess entropy---that
describe the intrinsic computational properties of the stacking
structures.

\end{abstract}

\pacs{
  61.72.Dd,   
  61.10.Nz,   
  61.43.-j,   
  81.30.Hd    
  \\
\begin{center}
  Santa Fe Institute Working Paper 03-02-XXX
  ~~~~~~~~arxiv.org/cond-mat/0302XXX
\end{center}
  }

\maketitle

\section{Introduction}
\label{Introduction}

In the first paper~\cite{vcc02b} of this two-part series we presented
a novel technique for the discovery and description of planar disorder in
\emph{close-packed structures} (CPSs):
\emph{\eM\ spectral reconstruction theory} (\eMSR\ or ``emissary'').
We showed that the technique allows one to build the unique, minimal, and
optimal model (an \eM) of a material's stacking structure from diffraction
spectra. In this sequel we demonstrate the technique using diffraction
spectra for disordered, polytypic zinc sulphide. Since the discovery of
\emph{polytypism} in mineral ZnS crystals by Frondel and
Palache~\cite{fp48,fp50} in 1948, much theoretical and experimental effort
has been expended to understand this phenomenon. (See, for example,
Steinberger,~\cite{s83} Mardix,~\cite{m86} and Sebastian and
Krishna.~\cite{sk94}) ZnS is an attractive system to study for a number of
reasons:

\begin{trivlist}

\item (i) {\it Simplicity of the unit cell and stacking rules.}
While many materials are known to be polytypic,~\cite{t91,sk94} the
constituent \emph{modular layers}~\cite{vc01} (MLs) can have a
complicated structure and complex stacking rules.~\cite{b80,t81,vc01}
For instance, in ideal micas there are more than a dozen atoms in a
unit cell, and there are six  
ways two MLs can be stacked. Kaolins and cronstedtites present
even more complexity.~\cite{vc01} In contrast, ZnS is simple in the
extreme: its basis is composed of but two atoms---a zinc and a
sulphur.~\cite{sk94} They are arranged in a double close-packed
hexagonal net, with one species displaced relative to the other by
a quarter body diagonal (as measured by the conventional unit cubic
cell) along the stacking direction.~\cite{sh90} We take this double
close-packed layer to be a ML.~\cite{vc01} The stacking of MLs
proceeds as for all CPSs; namely, there are three absolute orientations
each ML can occupy---call them $A$, $B$, and $C$---with the familiar stacking
constraint that no two adjacent MLs have the same orientation.~\cite{vc01}

\item (ii) {\it Complex polytypism.} ZnS is one of the most
polytypic substances known with over 185 identified crystalline
structures.~\cite{m86,t91,sk94} Of these, only about a dozen fairly
short-period polytypes (up to 21 MLs) are common in mineral ZnS,
with the remainder found in synthetic crystals. Some of the
crystal structures have repeat distances that extend over 100 MLs.
Also, many structures show considerable planar disorder. The wide
diversity of structural complexity remains one of the central mysteries
of polytypism.

\item (iii) {\it Solid-state transformations.} It is believed that there are
only two stable phases of ZnS, the low-temperature modification being the
$\beta$-ZnS or \emph{sphalerite} (3C~\cite{note2}) and the high-temperature
modification \emph{wurtzite} (2H) or $\alpha$-ZnS.~\cite{sk94}
The former transforms enantiotropically into the latter at 1024 C.
The plethora of structures suggests that most of them are not
in equilibrium but rather structures that are trapped in
a local minimum of the free energy and lack the necessary activation energy to
explore all of configuration space. It is possible to observe these
structures by annealing and then arresting the transformation upon
quenching. One can then study the various intermediate stages of
the transformation.

\item (iv) {\it Availability.} Polytypes of ZnS, both ordered and disordered,
are easily manufactured in the laboratory by a variety of
methods.~\cite{sk94} One of the more common is growth from the
vapor phase above temperatures of about 1100 C. Crystals can also
be grown from melt at high pressures, by use of chemical transport
and hydrothermally. The distribution of polytypes observed depends
on the method used.

\end{trivlist}

Nearly a dozen theories have been proposed to explain polytypism,
among them being the ANNNI model,~\cite{py84,y88} Jagodzinski's
disorder theory,~\cite{j54a} and Frank's screw dislocation
theory.~\cite{f51b} (For a complete discussion, see for example,
Verma and Krishna,~\cite{vk66} Trigunayat,~\cite{t91} and Sebastian
and Krishna.~\cite{sk94}) We will have little to say here about the
mechanisms that \emph{produce} various polytype structures. Instead,
our focus will be on describing the disordered structures so commonly seen.
We feel that an adequate description of the disordered structures---which
so far has been lacking---is warranted before one can evaluate models
that explain the formation of disorder and structures and, especially,
the solid-state phase transitions that lead to them.

Previous descriptions of planar disorder in single crystals of ZnS fall
into one of two categories: the \emph{fault model} (FM)~\cite{vcc02a,vcc02b}
and Jagodzinski's \emph{disorder model} (DM).~\cite{j49a,j49b,j49c,fjs86}
Applications of the FM include Roth's~\cite{r60} study of faulting
induced in hexagonal crystals grown from the vapor phase upon annealing.
Roth extracted correlation information from the diffraction spectra by Fourier
analysis and then derived analytical expressions relating how correlation functions 
decayed with both increasing separation between MLs and as a function of the fault
probability. He considered both randomly distributed growth and deformation
faults and found that for weakly disordered specimens deformation faulting
gave the best agreement with experiment.

Significant applications
of the FM to planar disorder in ZnS have been carried out by
Sebastian, Krishna, and coworkers. They studied the 2H-to-3C solid-state
transformation in vapor-grown ZnS crystals after annealing
between temperatures of 300 to 650~C.~\cite{spk82} By analyzing
and comparing the profiles of the integer-$l$ reflections~\cite{note4} to those
of the half-integer-$l$ reflections for weakly faulted crystals,
they found that the disorder was largely due to the random
insertion of deformation faults. They attributed slight discrepancies
between the observed and calculated profiles to the so-called
nonrandom insertion of faults. Sebastian and Krishna~\cite{sk84}
later studied the disordered stacking in 3C crystals grown from
the vapor phase, as well as those obtained from annealing 2H
crystals. They found that the structure of both the as-grown and
annealed crystals was best explained as randomly distributed twin
faults in the 3C structure. They concluded that the 2H-to-3C
transformation in ZnS proceeded by the nonrandom nucleation of
deformation faults occurring preferentially at two ML separations.

To better understand the nature of the nonrandom insertion of
deformation faults in the 2H structure, Sebastian and
Krishna~\cite{sk87a,sk87b} introduced a three-parameter model that
assigned separate probabilities to the random insertion of deformation
faults, as well as deformation faulting at two and three ML 
separations. They derived analytical expressions for the diffraction
spectra in terms of these parameters and concluded that both the
2H-to-3C and the 2H-to-6H transformations proceeded via the nonrandom
nucleation of deformation faults. Their analysis showed that these
transformations occurred simultaneously in different regions of
the same crystal. They attributed this to variations in the
stoichiometry. Sebastian~\cite{s88} gave a similar treatment that
came to the same conclusions. With the exception of Roth, all of
these analyses depended on carefully characterizing the change in
Bragg peaks as one introduces a small amount of disorder. We
have previously given a criticism of the FM approach
elsewhere.~\cite{vcc02a,vcc02b}

Jagodzinski's DM is a two-parameter model that assumes two thermodynamically
stable phases in CPSs:  the 2H and 3C. One therefore finds two
kinds of fault (and here we mean structure and not
mechanism): namely, cubic faults in the 2H and hexagonal faults in
the 3C. By choosing appropriate values of the two model parameters
one can also model 4H structures. Within this framework, an analytical
expression for the diffracted intensity is derived that depends on
the model parameters in a complicated manner. Nonetheless, one can select
model parameters that give the best agreement with experiment.

M\"{u}ller~\cite{m52} used this method to analyze faulted ZnS
diffraction spectra and found that while he was able to obtain
reasonable agreement between theory and experiment for a few spectra,
for many he was not.  Singer~\cite{sg63} re-examined this approach
and concluded that the DM applies when faulting is random, but when
the faulting is nonrandom, as many ZnS specimens are suspected to
be, the model fails. However, Frey {\it et al.}~\cite{fjs86}
studied the 3C-to-2H transformation in single crystals of ZnS using
the DM and were able to obtain excellent agreement between theory and
experiment. They fitted the experimental diffraction spectra to the DM's
analytical one. Due to the complicated nature of the expression, however,
they treated eight constants that depend on the two model parameters as
independent. From these eight fitted parameters they were able to find
the two model parameters that best fit each spectra.

One cannot help but raise questions concerning the mathematical rigor used
to find the model parameters in this way. We show elsewhere~\cite{vccup}
that the description of the stacking disorder as given by the DM is a special
constrained case of the $\Range =2$ computational mechanics approach. As in
the latter, in the DM there is no assumption of weak faulting and one does
use diffuse scattering to build the model. Since the spectra we analyze have
not been previously treated using the DM and it is a special case of our own,
we do not discuss the DM further here, but treat it in detail
elsewhere.~\cite{vccup}

A third possible method of discovering structural information about disordered
solids from diffraction spectra employs a reverse Monte Carlo (RMC)
technique.~\cite{pw98,wp98} In this method, one typically searches for a
configuration of constituent atoms such that a signal---{\it e.g.}, the
diffraction spectrum---estimated from the candidate structure most closely
matches the experimental signal. This technique can be applied
for the case of disorder in three dimensions. One drawback, however,
is that candidate structures are often found that are physically
implausible. One needs to impose assumptions to eliminate these.
To our knowledge, this technique has not been applied to polytypism,
and we do not consider it here.

In this work, we apply computational mechanics~\cite{cy89,fc98a,sc01}
to discover and describe disordered stacking sequences
in four previously published ZnS diffraction spectra.
We Fourier analyze each spectrum to find
correlation information between MLs and then calculate the probability
distribution of stacking sequences. From the latter, we reconstruct
the \eM\ that gives the stochastic process for the ML stacking and
compare it to previous FM analyses. From the reconstructed \eM, we
calculate the stacking entropy per layer, average stacking-fault
energy per Zn-S pair, memory length, hexagonality, and generalized
period.~\cite{vcc02b} We find that the diffraction spectra of the
four samples is well described using the computational mechanics approach.

We note that our primary purpose in the following is expository; that is,
we wish to demonstrate the efficacy of \eMSR\ on real data. 
Since we use diffraction spectra from older studies,~\cite{sk94} the analyses
given here are less than ideal. Specifically, we digitized data from
the published spectra and found that there was significant systematic error in each spectrum. 
Additionally, the experimental data was not reported with error bars.  
Despite these
possible shortcomings, \eMSR\ allows us to offer more
comprehensive analyses of the spectra than given by previous workers.

Our development is organized as follows. In \S \ref{Methods} we
outline our approach, including a brief discussion of the experimental
methods and our analysis. In \S \ref{Analysis}, we give the results of
\eM\ reconstruction for four experimental ZnS diffraction spectra and
contrast this to the FM approach when possible. In \S \ref{EMPhysics}
we calculate the stacking energies per Zn-S pair and the hexagonality
for the various structures from our reconstructed \eMs. In \S
\ref{Discussion} we give our conclusions and propose some directions for
future theoretical and experimental work.

\section{Methods}
\label{Methods}

The four diffraction spectra we analyze come from Sebastian and
Krishna~\cite{sk94} and are labeled SKXXX by the page (XXX) on which
they appear in that source. These data were collected in the 1980's
and since they no longer exist in numerical form,~\cite{seb01} we
digitized them from the diffractograms given in the 
Sebastian and Krishna publication.~\cite{sk94}
In this section, we give a brief synopsis of the experimental procedure,
discuss the assumptions made to analyze the data, and list the
corrections we apply to the experimental spectra.

\subsection{Experimental Details}

The experimental procedure is given in more detail
elsewhere.\cite{sk94,spk82,sk87b,s88} Briefly, the crystals were grown
from the vapor phase at a temperature in excess of 1100 C in the
presence of H$_2$S gas. Each crystal was needle-shaped, approximately
0.1 to 0.4 mm in diameter and 1 to 2 mm in length. Two of the four 
crystals were further annealed for one hour at 
300 and 500~C. These investigations were performed to better
understand the fault structures they contain, as well as study
the solid-state transformations that ZnS crystals undergo.

The intensity along the $10.l$ reciprocal lattice row was recorded
using a four-circle single-crystal diffractometer for each specimen
in steps of approximately $\Delta l = 0.005$. (Our definition of
$l$ differs slightly from that of Sebastian and Krishna,~\cite{sk94}
so the $l$-increment we report also differs.) The crystal and the
counter were held stationary while the crystal was illuminated with
Mo$K_{\alpha}$ radiation. The sharp reflections along the $h - k
= 0 \mod 3 $ rows were used to orient each crystal. The divergence
of the incident beam was adjusted to cover the mosaic spread for
each crystal. The experimental diffraction spectrum is reported as
the total number of counts versus $l$. The crystals were examined
under a vickers projection microscope and did not show signs
of kinking or shearing, even after annealing. They did show parallel
striations or stripes perpendicular to the stacking direction.

\subsection{Assumptions}

To make the analysis tractable, we employ the following assumptions common
in the analysis of planar disorder in ZnS:

\begin{trivlist}

\item (i) {\it Each ML is perfect and free of distortions and
defects.} We assume that each ML is identical and the MLs themselves
are undefected. That is, each ML is crystalline in the strict
sense, with no point defects, impurities, or distortions in the
two-dimensional lattice structure. This clearly precludes the possibility
of screw dislocations, which are known to play a role in the polytypism
of some ZnS crystals.~\cite{sk94,m88} Since each of the crystals
we analyze was examined under a vickers projection microscope and
no such dislocations were seen, and the crystals retained their
shape after annealing, this seems reasonable. (It is known~\cite{sk94}
that during solid-state transformations of specimens of ZnS with
an axial screw dislocation the specimen will exhibit "kinking" with
a characteristic angle of $19^0 28'$.)

\item (ii) {\it The spacing between MLs is independent of the
local stacking arrangement.} There is known to be some slight
dependence of the inter-ML spacing depending on the
polytype.~\cite{s88,sk94} For the 2H structure in ZnS, the inter-ML
spacing is measured to be 3.117 \AA.  For the 3C structure, the
cubic cell dimension is $a_c = 5.412$ \AA\ which gives a corresponding
inter-ML spacing of 3.125 \AA.  Therefore, to an excellent
approximation, this spacing is independent of the local stacking
environment in ZnS.

\item (iii) {\it The scattering power of each ML is the same.}
We assume that each ML diffracts x-rays with the same intensity.
There is no reason to believe that this is not so, unless absorption
effects are important or the geometry of the crystal is such that
each ML does not have the same cross-sectional area.

\item (iv) {\it The stacking faults extend over the entire fault
plane.} Examination under microscope indicates that this is generally
true. However, Akizuki~\cite{a81} found evidence that the faults
do not extend completely over the faulted MLs by examining a
partially transformed ZnS crystal under an electron microscope.

\item (v) {\it We assume that the ``stacking process" is stationary.}
We simply mean that the faults are uniformly distributed though
out the crystal. Put another way, we assume that probability of
finding a particular stacking sequence is independent of its location
in the crystal. This does not, however, preclude regions of crystal
structure interspersed between regions of disorder. It simply means
that the statistics of the stacking does not change as one moves
from one end of the crystal to the other.

\end{trivlist}

Notably absent from this list are any assumptions about the crystal
structure present (if any) and how the sample might deviate from
that structure. In contrast to the FM, we invoke no {\it a priori} 
structural assumptions concerning the stacking sequence.

\subsection{Corrections to the Experimental Spectra}

We corrected each spectrum for the following effects:

\begin{trivlist}

\item (i) {\it The atomic scattering factor.} This correction
accounts for the spatial distribution of electrons, as well as for
the wavelength of the incident radiation and angle of reflection.
Calculations of these effects are given in standard tables~\cite{hws92}
and we employ them in our work.

\item (ii) {\it The structure factor.} The structure factor~\cite{sk94}
accounts for the two-atom basis in ZnS.

\item (iii) {\it Anomalous scattering factors.} Also called dispersion
factors, the anomalous scattering factors correct for the binding
energy of the electrons and the phase shifts.~\cite{hws92} For our
case, we find these to be small, but have included them nonetheless.

\item (iv) {\it Polarization factor.} We use the standard correction
factor for unpolarized radiation.~\cite{w69}

\end{trivlist}

Factors we did {\em not} correct for include the following:

\begin{trivlist}

\item (i) {\it Thermal factors.} At room temperature, this effect
is small for ZnS.~\cite{v01}

\item (ii) {\it Absorption factor.} For the geometry of the ZnS
crystals we analyze, the linear mass coefficient~\cite{w97,m73} is
much larger than the thickness, therefore we ignore it.

\item (iii) {\it Instrument resolution.} This is not reported with the 
experimental data, so we do not deconvolve the spectrum.

\end{trivlist}

\section{Analysis}
\label{Analysis}

We now give the structural analysis of four experimental diffraction spectra
taken from Sebastian and Krishna.~\cite{sk94} We apply \eMSR\ to each to
build a model that describes the stacking process. From our model, we
calculate various measures of intrinsic computation for each spectra. 
We compare our results with that obtained by previous researchers using the
FM. Since the experimental spectra are not reported with error bars, we are
unable to set an error threshold $\Gamma$ as required in \eMSR. Instead,
we found that each increase in the memory length $\Range$ continues 
to give a better description of each spectra.~\cite{v01} We perform 
\eM\ reconstruction up to $\Range = 3$ for each spectra. To find the CFs
from the \eMs, we take a sample length 400 000, as generated by the \eM. 
The diffraction spectra are calculated using 10 000 MLs. The experimental
spectra are normalized to unity over the $l$-interval used for reconstruction,
as are the spectra calculated from each \eM. For the spectra calculated from
the FM, we set the overall scale to best describe the Bragg peaks as shown in
Sebastian and Krishna.~\cite{sk94} We also calculate the profile
${\cal R}$-factor~\cite{vcc02b} to evaluate the agreement between experiment
and theory for each spectrum.  


\subsection{SK134}

\begin{figure}
\begin{center}
\resizebox{!}{5cm}{\includegraphics{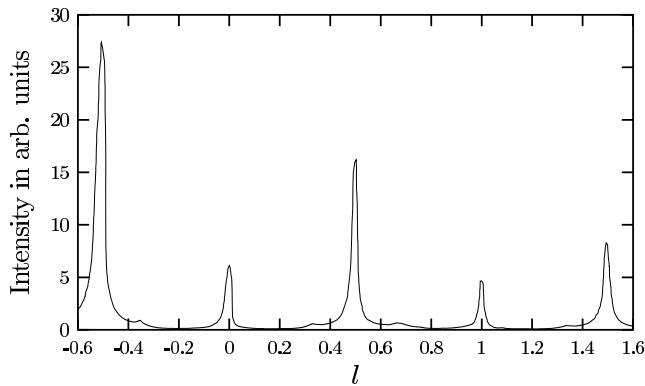}}
\end{center}
\caption{
  Diffraction spectrum along the $10.l$ row from SK134 corrected for atomic scattering factors,
  the structure factor, dispersion factors, and the polarization
  factor.~\cite{hws92,w97,w69} We see that the spectrum is not periodic in
  $l$, as it should be. This indicates that there exist significant errors
  in the data.
  }
\label{fig:dpSK134}
\end{figure}

The corrected diffraction spectrum along the $10.l$ row for an as-grown
2H ZnS crystal is shown in Fig.~\ref{fig:dpSK134}. One immediately notices
that the spectrum is not periodic in $l$, as it should be, but instead
suffers from variations in the intensity. The peaks at $l= 0.0$ and
$l = 1.0$ are of similar intensity, but the peaks at $l = -0.5$, 0.5 and 1.5 
seem to suffer from a steady decline in intensity.
We can therefore be confident that this spectrum
contains substantial systematic error as reported by the experimentalists.

This difference in diffracted intensity between peaks results from
the finite thickness of the Ewald sphere due to the divergence of
the incident beam.~\cite{sk87b,sk94} A suitable choice of geometry
can minimize, but not eliminate these effects, such that one finds
only a gradual variation in $I$ with $\Delta l$.
Since analysis by
the FM depends only on the change of the shape and the position of
the Bragg-like peaks, such a slow variation of the diffracted
intensity with $l$ will not affect the conclusions drawn from an
FM analysis. It is possible to correct these effects,~\cite{pplg87,sk94}
but this has not been done in the literature.

\begin{figure}
\begin{center}
\resizebox{!}{5cm}{\includegraphics{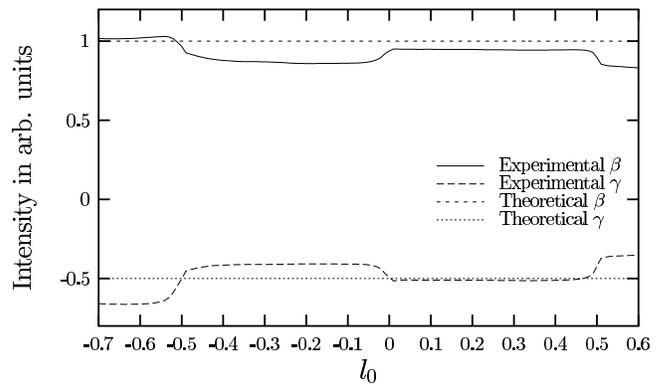}}
\end{center}
\caption
  {
  Experimental and theoretical figures-of-merit---$\beta$ and $\gamma$---as
  a function of $l_0$ for diffraction spectrum SK134. We define $l_0$ as
  the point at which integration over a unit $l$-interval is initiated. We
  see that $l_0 \approx 0.04$ gives the best agreement with the theoretical
  values.
  }
\label{fig:SK134.1.cor.ext}
\end{figure}

Our analysis depends on selecting an appropriate $l$-interval
where variations in diffracted intensity due to these experimental effects
are minimized. It is important, then, to select an interval that is relatively
error-free. As discussed in Part I,\cite{vcc02b} there are two criteria,
called \emph{figures-of-merit} and denoted $\gamma$ and $\beta$, one can
use for this. It can be shown that in an error-free spectrum the
parameters must be equal to constant values $-1/2$ and $1$, respectively,
for \emph{any} unit $l$-interval,
regardless of the amount of planar disorder present. The extent that
$\beta$ and $\gamma$ differ from their theoretical values over a given
$l$-interval measures how well the diffraction spectrum over the
interval can be represented by a physical stacking of MLs. It makes sense,
then, to choose an $l$-interval for \eM\ reconstruction such that the
theoretical values of the figures-of-merit are most closely realized.
This does not, of course, guarantee that the interval is error-free.
Glancing at
Fig.~\ref{fig:SK134.1.cor.ext} shows that $\gamma = -0.51$ and $\beta = 0.95$
over the interval $l \in [0.04, 1.04]$.

\begin{figure}
\begin{center}
\resizebox{!}{5cm}{\includegraphics{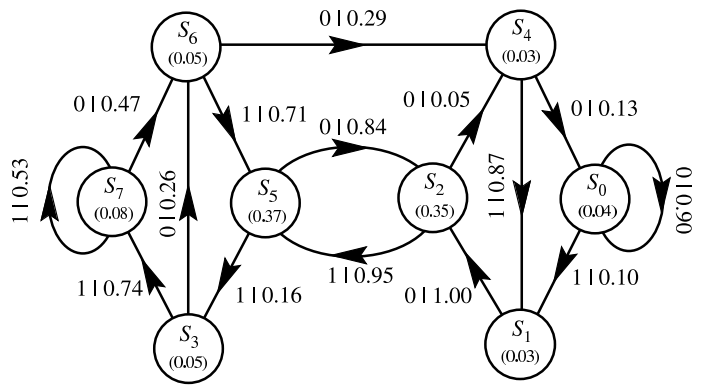}}
\end{center}
\caption{
The $\Range =3$ \eM\ reconstructed for SK134. The recurrent causal states
  $\{\CausalState_0 \ldots \CausalState_7\}$ are labeled (in decimal notation)
  by the last three (binary) spins seen and the asymptotic state probabilities
  are given for each state. Edge labels $s|p$ indicate a transition on
  spin $s$ with probability $p$. The large asymptotic state probabilities
  for the $\CausalState_2$ and $\CausalState_5$ states, as well as the large
  state-transition probabilities between them, show that this is predominately
  a 2H crystal with some faulting.
  }
\label{fig:dB3.134}
\end{figure}

We perform \eMSR~\cite{vcc02b} and find that the smallest-$r$
\eM\ that gives reasonable agreement between the measured and \eM\ spectra
has a memory length of $\Range =3$. The reconstructed \eM\ is shown in
Fig.~\ref{fig:dB3.134}. The large asymptotic state probabilities for the
$\CausalState_2$ and $\CausalState_5$ causal states (CSs), as well as the
large inter-state transition probabilities between them, indicate this is
predominantly a 2H crystal. More specifically, the probability of seeing
sequences 1010 and 0101, corresponding to the 2H cycle, have a combined total
weight of about $64\%$. The remaining probability is distributed among the
other fourteen length-$4$ sequences. It is tempting to interpret the remaining
structure in terms of faults and, indeed, it seems we can.

Let us treat the transitions $s = 0$ from $\CausalState_4$ and $s = 1$ from
$\CausalState_0$ as though they are missing for the purposes of a fault
analysis. This implies that the sequences $0001$ and $1000$ are disallowed.
Of course, this cannot be exactly true, as the CS $\CausalState_0$ would then
be isolated from the rest of the \eM. In this case, we would say that the
\eM\ is not {\em strongly connected} and, as such, cannot represent a physical
stacking of MLs. However, the combined probability weight of these two
sequences is $ < 1\%$, so neglecting them gives only a small error in our
intuitive understanding of the faulting structure.

Then on the left half of the \eM\ there is structure associated with a 2H
deformation fault $[\CausalState_3\CausalState_7\CausalState_6\CausalState_5]$
with probability weight $0.16=\Prob(1011)+\Prob(0111)+\Prob(1110)+\Prob(1101)$.
We can interpret the causal-state cycle (CSC) 
$[\CausalState_3\CausalState_6\CausalState_4\CausalState_1\CausalState_2\CausalState_5]$
as a layer-displacement fault and see that it is assigned a probability weight
of $0.06$. The right portion contains the CSC
$[\CausalState_4\CausalState_1\CausalState_2]$ with probability weight $0.06$,  
which is associated with growth faults. The CSCs $[\CausalState_0]$ and
$[\CausalState_7]$, identified as 3C structure, have a combined weight of
$0.08$.

Given these observations, a possible interpretation suggested by
the \eM\ is that SK134 has crystal structures and faults in the proportions
given in Table \ref{Table:SK134}. The decomposition there is sensible since
there is an underlying crystal structure present and the smaller, faulting
paths are not too large. As we will see, these need not always be the case.
Sebastian and Krishna~\cite{sk94} have analyzed this diffraction
spectrum using the FM and found that approximately one in every
twenty MLs is deformation faulted, so they described the stacking
structure as a faulted 2H crystal with 5\% random deformation
faulting. This is equivalent to assigning CSCs 
responsible for deformation faulting a total probability weight of 0.17.
We compare the structure analyses of the two models in 
Table \ref{Table:SK134}.

\begin{table}
\begin{center}
\begin{tabular}{lrr}
\hline
\hline
 Structure                         & \EM\      & Fault Model \\
 \hline
 \textrm{2H}                       &  $ 64\% $ & $83\%$      \\
 \textrm{3C}                       &  $  8\% $ & $ 0\%$      \\
 \textrm{Deformation fault}        &  $ 16\% $ & $17\%$      \\
 \textrm{Growth fault}             &  $  6\% $ & $ 0\%$      \\
 \textrm{Layer-displacement fault} &  $  6\% $ & $ 0\%$      \\
\hline
\hline
\end{tabular}
\end{center}
\caption{Structural decomposition of SK134 according to the reconstructed
  \eM\ of Fig.~\ref{fig:dB3.134} and according to the fault model analysis of
  Sebastian and Krishna.~\cite{sk94} The latter is valid only under the
  assumption of weak faulting.
  }
\label{Table:SK134}
\end{table}

We see that both analyses agree that the dominant structure is 2H, though the
\eM\ attributes less of the crystal structure to this ``parent" phase.
Similarly, both find that structures associated with deformation faulting
are important and assign them almost equal weights.

They differ, however, in that the \eM\ finds additional faulting structures
(growth and layer-displacement faulting), as well as 3C crystal structure.
Since this crystal, if annealed at sufficient temperatures for a long enough
time, will transform into a twinned 3C structure, the latter is easily
understood as nascent structure in that process. Finding the presence of
weak 3C structure is not unreasonable since there is some slight enhancement
of the diffracted intensity at $l \approx 1/3$ and $l \approx 2/3$. The other
faulting structures seen are less easily understood. Growth faults, so-named
because they primarily are formed during the growth of the 2H crystal, are
not expected to play an important role in solid-state transformations of
ZnS.~\cite{r60} Their presence here may result from the
initial growth of the crystal. The small amount of layer-displacement
structure could be seen as two adjacent, yet oppositely
oriented deformation faults. That is, a deformation fault
in a 2H structure is simply a spin flip in the H\"{a}gg
notation,~\cite{vcc02b} so that a sequence $ \ldots 010101 \ldots
$ transforms to $ \ldots 01\underline{1}101 \ldots $ as a result
of one deformation and then to $ \ldots 011\underline{0}01 \ldots
$ upon another resulting in a layer-displacement fault $ \ldots
01\underline{10}01 \ldots $. This might imply some coordination
between faults. Or, the mechanism of layer-displacement faulting
may play some minor role in the solid-state transformation.

However, one cannot disambiguate these from the available spectra
and the reconstructed \eM. The \eM\ only provides information
about the structure. We must look outside the \eM\ to formulate
an understanding of \emph{how} the polytype came to be stacked
in this way. This is the critical difference between faulting
\emph{mechanism} and faulting \emph{structure}. In the former,
a physical process is responsible for causing the MLs to shift
or deviate from a perfect crystal. In the latter, in the limit of
weak faulting, the physical process leads to a given (statistical)
structure. In this limit it may be possible to postulate with some
certainty that the mechanism resulted in the observed structure. For
more heavily faulted crystals, however, such an identification of
structure with mechanism is dubious. Other techniques, such as numerical
simulations~\cite{kp88,e90,stkp96,sp96,sp96a,sp97,g00,g01} or analysis
of a series of crystals in various stages of the transformation, are
necessary to unambiguously determine the mechanism.

\begin{figure}
\begin{center}
\resizebox{!}{5cm}{\includegraphics{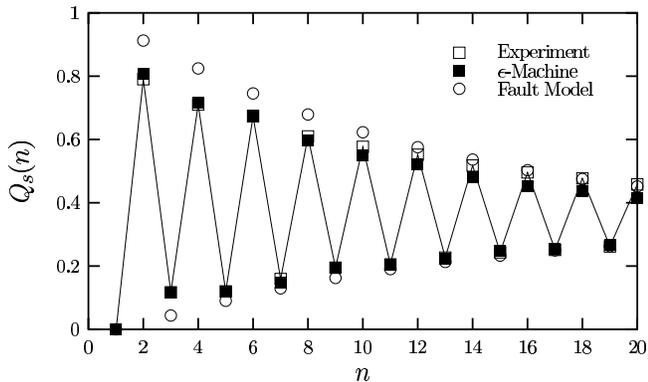}}
\end{center}
\caption[.]{$Q_s(n)$ versus $n$ for experimental spectrum SK134 (open squares
  connected by solid line), the fault model (open circles), and the $\Range =3$
  \eM\ (solid squares). The $Q_s(n)$ are defined
  only for integer values of $n$, but lines are drawn connecting adjacent
  points as an aid to the eye. We see good agreement up to $n \approx 15$,
  after which the $\Range = 3$ approximate correlation functions die too
  quickly to their asymptotic value of $1/3$.
  }
\label{fig:qsSK134.r.3.5a}
\end{figure}

\begin{figure}
\begin{center}
\resizebox{!}{5cm}{\includegraphics{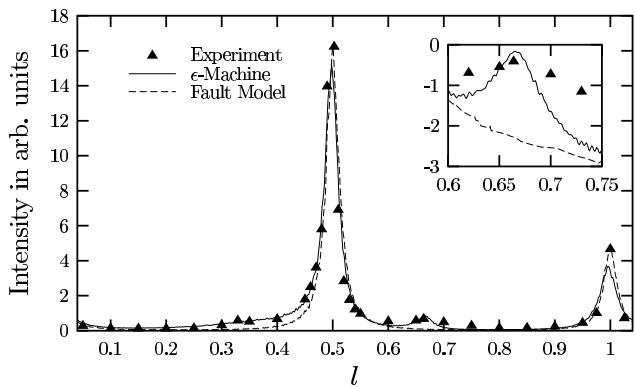}}
\end{center}
\caption{Comparison of the experimental diffraction spectrum SK134 along the
  $10.l$ row (triangles) for a disordered ZnS single crystal with a spectra
  estimated from the FM with $5\%$ deformation faulting (dashed line) and
  $\Range =3$ \eM\ (solid line). The vertical scale in the inset is logarithmic
  intensity. For clarity, we
  report only a few representative data points from the experimental
  diffraction spectrum. In our analysis, however, we used the much finer mesh
  reported in the experimental data.
  We find that the \eMSR\ gives ${\cal R}_{\epsilon M} = 20\%$, while the
  FM gives ${\cal R}_{\mathrm {FM}} = 33\%$.
  }
\label{fig:SK134.1a1.cor.ext}
\end{figure}

Returning to the analysis of SK134, Fig.~\ref{fig:qsSK134.r.3.5a} compares
the CFs obtained from the experimental diffraction spectrum, those
obtained from the \eM\ and those from the FM. There is reasonable agreement
between the experimental and
\eM-predicted CFs. For small $n$, however, the FM overestimates the
amplitude in the oscillations in $Q_s(n)$. The experimental diffraction
spectrum is compared to that calculated from the FM and \eM\ in
Fig.~\ref{fig:SK134.1a1.cor.ext}. Both models give good agreement
near the Bragg peaks at $l= 0.5$ and $l = 1.0$, with perhaps the FM
performing a little better at $l= 1.0$. The diffuse scattering near the
shoulders of the $l= 0.5$ peak are better represented by the \eM. We calculate 
the profile ${\cal R}$-factor between experiment and the FM to be
${\cal R}_{\mathrm {FM}} = 33\%$ and between experiment and 
\eMSR\ to be ${\cal R}_{\epsilon M} = 20\%$.    

From the \eM\ we calculate the three length parameters to be
$\Range =3$ ML, $\mathcal{P} = 4.8$ ML, and $\lambda_c = 9.5 \pm 0.5$ ML. 
The three measures of intrinsic computation are found to be 
$\hmu = 0.50$ bits/ML, $\Cmu = 2.3$ bits, and $\EE = 0.75$ bits.   
We return to discuss and compare these for all of the samples in a
later section.

\subsection{SK135}

\begin{figure}
\begin{center}
\resizebox{!}{5cm}{\includegraphics{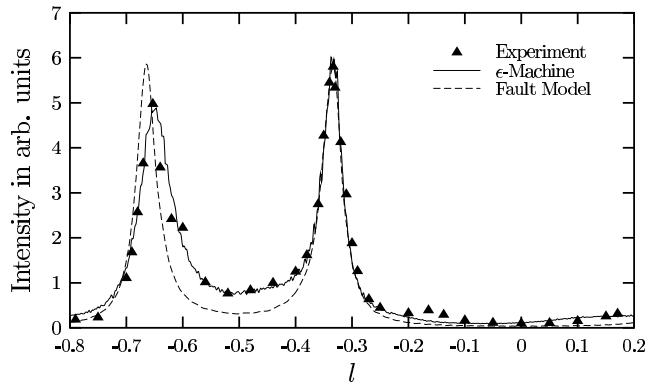}}
\end{center}
\caption[.]{Comparison of the experimental diffraction spectrum SK135
  along the $10.l$ row (triangles) for a disordered 3C single crystal
  with the diffraction spectra calculated from the FM with $12\%$
  twinned faulting (dashed line) and the \eM\ (solid line).
  We find that the \eMSR\ gives ${\cal R}_{\epsilon M} = 13\%$, while the
  FM gives ${\cal R}_{\mathrm {FM}} = 33\%$.
  }
\label{fig:SK135.1a1.cor.ext}
\end{figure}

\begin{figure}
\begin{center}
\resizebox{!}{5cm}{\includegraphics{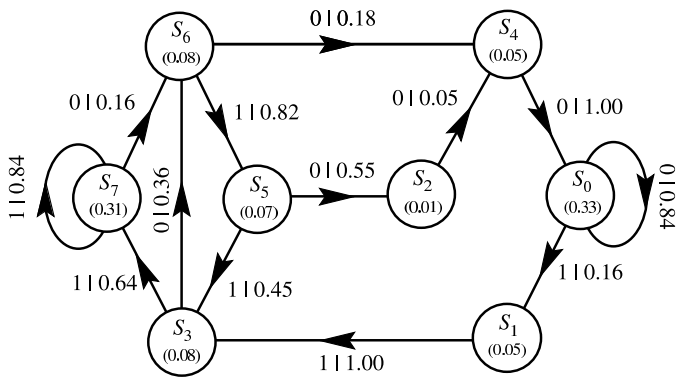}}
\end{center}
\caption{The reconstructed $\Range =3$ \eM\ for SK135. The strong self-loop
  transition probabilities between causal states $\CausalState_0$ and
  $\CausalState_7$ as well as their large asymptotic state probabilities
  indicate that the $\ldots 0000 \ldots$ and $\ldots 1111 \ldots$  structures
  are important. This, then, is a twinned 3C crystal.
  }
\label{fig:dB3.135}
\end{figure}

The next sample we examine is a twinned 3C crystal obtained by
annealing a 2H crystal at 500~C for 1 h. The diffraction spectrum
for this crystal is given in Fig.~\ref{fig:SK135.1a1.cor.ext}. We
find that figures-of-merit are closest to their theoretical values
over the interval $l \in [-0.80,0.20]$ with values $\beta = -0.50$
and $\gamma = 0.93$. The smallest-$r$ \eM\ that gives reasonable
agreement with experiment was found at $\Range =3$ and has a
profile ${\cal R}$-factor of ${\cal R}_{\epsilon M} = 13\%$.
The resulting \eM\ is shown in Fig.~\ref{fig:dB3.135}. 
Based on the presence of asymmetrically
broadened peaks and the absence of peak shifts, a FM analysis~\cite{sk94}
finds this sample to be a twinned 3C crystal with $12\%$ twinned faulting.
The profile ${\cal R}$-factor between experiment and the FM is found 
to be ${\cal R}_{FM} = 33\%$.

The large CS probabilities associated with CSs $\CausalState_0$ and
$\CausalState_7$, as well as their large self-loop transition probabilities,
suggest that this is a twinned 3C crystal. We also note that the transitions
corresponding to antiferromagnetic paths ($0101$ and $1010$) have a
relatively small combined weight of only about $4\%$. In fact, the probability
weight for the $0101$ path is zero. (The transition from $\CausalState_2$
to $\CausalState_5$ is missing.) This indicates that 2H structure has
largely been eliminated. In
addition to the $0101$ path, the $1001$ and $0010$ paths are also missing.
This implies that twinned faulting is important, but also the
remnant of the $1010$ path has some role. Instead of a simple twinned 
fault $[\CausalState_7\CausalState_6\CausalState_4\CausalState_0]$ giving
the sequence $\ldots 1111|0000 \ldots$, where the vertical line indicates
the fault plane, the path
$[\CausalState_7\CausalState_6\CausalState_5\CausalState_2\CausalState_5\CausalState_0]$
giving the sequence $\ldots 1111|01000 \ldots$ has approximately twice as much
probability weight associated with it. In the \eM's right portion twinned
faulting $[\CausalState_0\CausalState_1\CausalState_3\CausalState_7]$ is
largely responsible for the $(0)^*$ 3C cycle converting to the $(1)^*$ 3C
cycle and we also observe that double deformation faulting
$[\CausalState_0\CausalState_1\CausalState_3\CausalState_6\CausalState_4]$
plays a role.

It is interesting to mention that, while a ML of ZnS has spin-inversion
symmetry~\cite{vc01} and, thus, the one-dimensional Hamiltonian describing
the energetics of the stacking is also spin invariant, in general \eMs\ need
not be spin-inversion invariant. [Note that the \eM\ in Fig.~\ref{fig:dB3.135}
is \emph{not} spin-inversion invariant.] There is, of course, no reason why we
should demand spin-inversion invariance. After all, then one could never have
a crystal of purely one 3C structure or the other. However, since this crystal
was initially in the 2H structure---which is spin-inversion invariant---it is
curious that this is not preserved as the crystal is annealed. That is, there
is no reason to expect that faulting should occur preferentially with one
chirality. Notably, the FM \emph{always} assumes spin-inversion symmetry.  

\begin{figure}
\begin{center}
\resizebox{!}{5cm}{\includegraphics{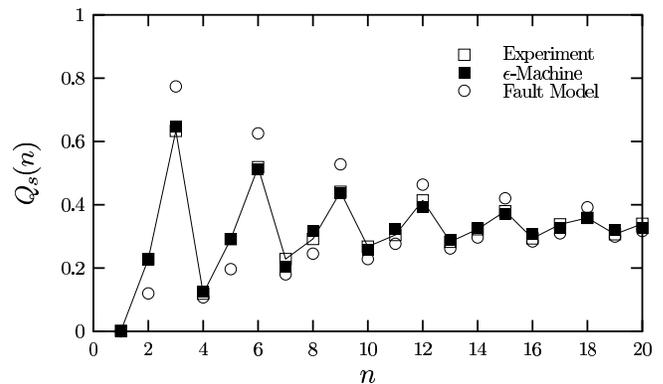}}
\end{center}
\caption{Comparison of the $Q_s(n)$ versus $n$ for experimental spectrum SK135
  (open squares), the \eM\ (solid squares), and the FM (open circles). The
  \eM\ gives good agreement with experiment, while the FM overestimates the
  oscillation magnitude.
  }
\label{fig:qsSK135.r.3a}
\end{figure}

Examining the $Q_s(n)$ estimated from experiment with those found from the
\eM\ in Fig.~\ref{fig:qsSK135.r.3a}, we find reasonable agreement up to
$n \approx 20$.  The $Q_s(n)$ found from the FM generally overstate the
magnitude of the oscillations in the CFs.

We can further examine the diffraction spectra. In
Fig.~\ref{fig:SK135.1a1.cor.ext}, the diffraction spectrum found from the
FM and the \eM\ are compared with experiment. The \eM\ gives a good fit,
except perhaps at a shoulder in the experimental spectrum at $l = -0.6$
and the small rise at $l = -0.16$. The latter might be understood as a
minor competition between the 3C and 6H CSCs that is not being well modeled
at $\Range = 3$. Comparison of the diffraction spectrum from the FM with that from
experiment reveals good agreement with the peak at $l= -0.33$ and poor
agreement with the one at $l=-0.67$. This is not surprising as the FM did
not use the peak at $l=-0.67$ to find the faulting structure. Likewise,
the diffuse scattering between peaks is not at all well represented by the
FM. Additionally, the small rise in diffuse scattering at $l = -0.16$ is
likewise absent in the FM diffraction spectrum.

From the \eM\ we calculate the three length parameters to be
$\Range =3$ ML, $\mathcal{P} = 5.6$ ML, and $\lambda_c = 4.4 \pm 0.7$ ML.
The three measures of intrinsic computation are found to be
$\hmu = 0.59$ bits/ML, $\Cmu = 2.5$ bits, and $\EE = 0.71$ bits.

\subsection{SK137}

\begin{figure}
\begin{center}
\resizebox{!}{5cm}{\includegraphics{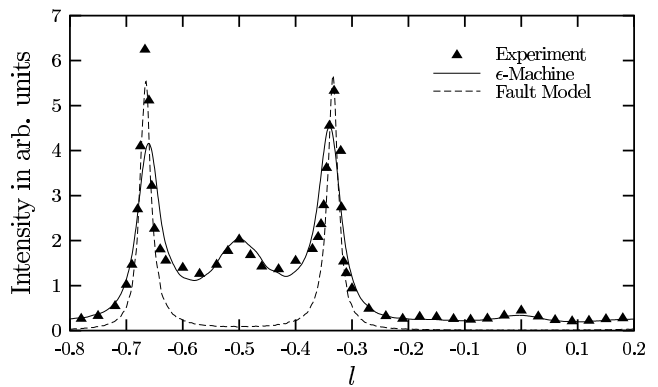}}
\end{center}
\caption{Comparison of the experimental diffraction spectrum SK137 along the
  $10.l$ row (triangles) for a disordered 3C single crystal with the
  diffraction spectra calculated from the FM with $6.8\%$ twinned faulting
  (dashed line) and the \eM\ (solid line). The profile ${\cal R}$-factor 
  between experiment and the \eM\ calculated diffraction pattern is
  ${\cal R}_{\epsilon M} = 17\%$. The FM gives considerably worse agreement,
  with a calculated profile ${\cal R}$-factor of ${\cal R}_{FM} = 58\%$ between
  it and experiment.
  }
\label{fig:SK137.1a1.cor.ext}
\end{figure}

\begin{figure}
\begin{center}
\resizebox{!}{5cm}{\includegraphics{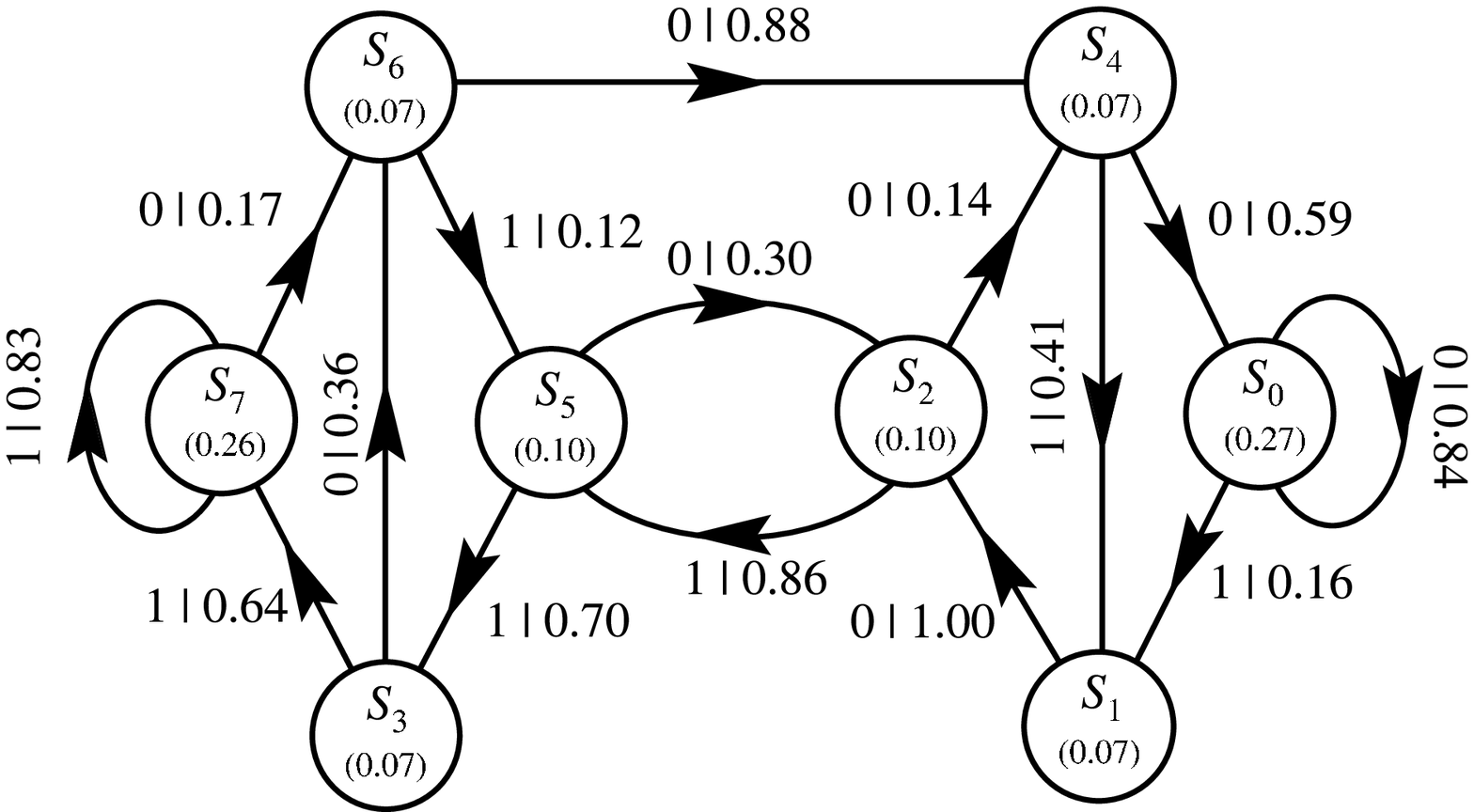}}
\end{center}
\caption{The reconstructed $\Range =3$ \eM\ for SK137. The strong self-loop
  transition probabilities between causal states $\CausalState_0$ and
  $\CausalState_7$, as well as their large asymptotic state probabilities,
  suggest that the $\ldots 0000 \ldots$ and $\ldots 1111 \ldots$ structures
  are important. Notice that, unlike the \eM\ for SK135, the CSC
  $[\CausalState_2\CausalState_5]$ is present, suggesting that associated
  2H structure is present. The absence of the transition between CSs
  $\CausalState_1$ and $\CausalState_3$ implies that the $0011$ sequence, and
  therefore the CSC associated with the 6H structure, is not present.
  }
\label{fig:dB3.137}
\end{figure}

The third experimental spectrum we analyze comes from an as-grown
disordered, twinned 3C crystal. The diffraction spectrum for this
crystal along the $10.l$ row is shown in Fig.~\ref{fig:SK137.1a1.cor.ext}.
The figures-of-merit are closest to their theoretical values over
the interval $l \in [-0.8,0.2]$ with values of $\gamma = -0.49$
and $\beta = 0.98$. We performed \eM\ reconstruction up to $\Range =3$
and found that this produces reasonable agreement with experiment giving
a profile ${\cal R}$-factor of ${\cal R}_{\epsilon M} = 17\%$. The
$\Range =3$ approximate \eM\ in shown in Fig.~\ref{fig:dB3.137}.
A FM analysis finds SK137 to be a twinned 3C crystal with $6.8\%$ twinned
faulting.~\cite{sk94} The FM-calculated diffraction spectrum has a 
profile ${\cal R}$-factor of ${\cal R}_{\epsilon M} = 58\%$ when compared 
with experiment.  

\begin{figure}
\begin{center}
\resizebox{!}{5cm}{\includegraphics{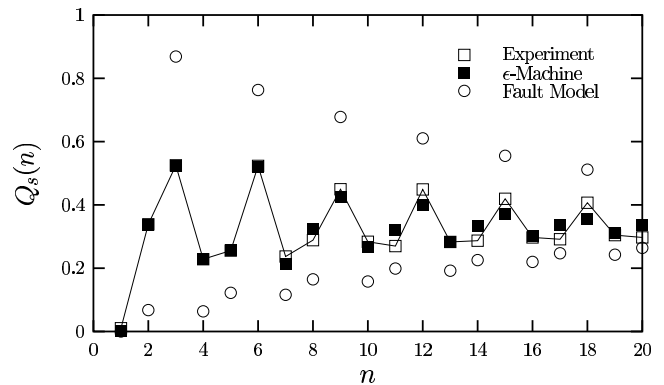}}
\end{center}
\caption{Comparison of the $Q_s(n)$ versus $n$ for experimental spectrum
  SK137 (open squares), the $\Range =3$ approximate \eM\ (solid squares),
  and the FM (open circles).  
  }
\label{fig:qsSK137.r.3a}
\end{figure}

A comparison of the CFs from experiment, the FM, and the \eM\ is shown in
Fig.~\ref{fig:qsSK137.r.3a}. For smaller $n$, the \eM\ gives good agreement
with experiment, although the error increases at larger $n$. As shown
elsewhere,~\cite{v01} the experimental CFs maintain small, but persistent
oscillations about their asymptotic value of $1/3$ up to $n \approx 40$,
while the CFs derived from the \eM\ effectively reach this asymptotic value
at $n \approx 25$. This leads us to speculate that there is some
structure in the stacking process that the \eM\ is missing. We expect that
reconstruction at $\Range =4$ will prove interesting here. This has not yet
been completed.

The FM fares markedly worse. It substantially overestimates the strength
of the oscillations in the CFs for all $n$.

A comparison of the diffraction spectra for experiment, the FM, and the
\eM\ is given in Fig.~\ref{fig:SK137.1a1.cor.ext}. The \eM\ gives reasonable
agreement everywhere except around the Bragg peaks at $l = -0.33$ and
$l = -0.67$. Here the \eM\ gives a value for the peak intensity $15\%$ and
$35\%$ lower, respectively, than experiment. The FM does much better at the
Bragg peaks, as one might expect. The diffuse scattering between the peaks,
and especially the broad-band rise in intensity near $l=-0.5$, are simply
missing in the FM diffraction spectrum, however. The \eM\ fit in this region
is substantially better, picking up a number of important spectral features,
such as broadband components and broadened peaks.

What does the \eM\ imply about the stacking process? All CSs and allowed
transitions are present except for the transition between $\CausalState_1$
and $\CausalState_3$. This absent transition implies that the $0011$ stacking
sequence is not present in SK137. This, then, means that the $000111$ sequence,
and hence the CSC
$[\CausalState_7\CausalState_6\CausalState_4\CausalState_0\CausalState_1\CausalState_3]$
associated with the 6H structure,\cite{vcc02b} is also absent. Therefore,
in this twinned 3C crystal there is no 6H structure.
This is surprising, since many ZnS spectra show enhancement about the 6H
positions during solid-state phase transitions from 2H to 3C. In
Fig.~\ref{fig:SK137.1a1.cor.ext} there is (arguably) a slight increase
in diffracted intensity at $l = -0.16$ and $l = 0.16$. So the absence of the
6H structure does seem echoed in the experimental spectrum. There is, however,
a large broadband increase in intensity about $\l=-0.5$ and a much smaller
increase about $\l = 0.0$. Reflections at these $l$ are usually associated
with 2H structure, with the half-integer peaks carrying three times the
intensity of the integer peaks. The \eM\ does show that the CSC
$[\CausalState_2\CausalState_5]$ associated with the 2H structure is present.
The frequency of occurrence of the 0101 and 1010 stacking sequences together
make up about $12\%$ of the total probability weight on the \eM. Even though 
$P_{CSC}([\CausalState_2\CausalState_5]) \ll 1$,  
it is not unreasonable to suggest that 2H structure is present.

As with the other spectra, we can calculate several characteristic length 
and information- and computational-theoretic quantities from the \eM.  
We find the three length parameters to be
$\Range =3$ ML, $\mathcal{P} = 6.7$ ML, and $\lambda_c = 12 \pm 3$ ML.
The three measures of intrinsic computation are found to be
$\hmu = 0.65$ bits, $\Cmu = 2.7$ bits, and $\EE = 0.79$ bits.


\subsection{SK229}

\begin{figure}
\begin{center}
\resizebox{!}{5cm}{\includegraphics{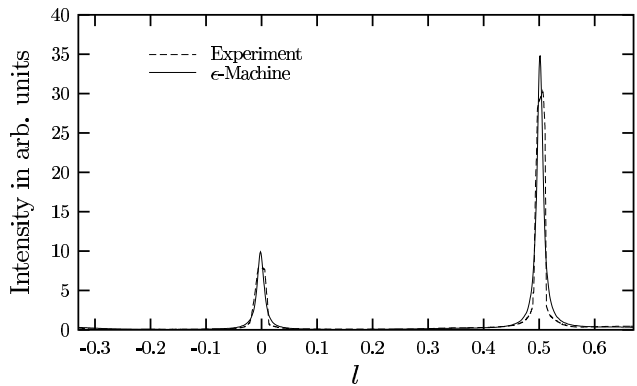}}
\end{center}
\caption{A comparison of the experimental diffraction spectrum 
  SK229 (dashed line) and that calculated from the reconstructed \eM\ (solid line).
  There is generally good agreement between two, except that the Bragg-like peaks 
  from the \eM\ are slightly displaced from the experimental spectra
  and the maximum from the \eM\ overestimates experiment. The \eM\ also 
  has some difficulty reproducing the shape of the experimental spectra. 
  The profile ${\cal R}$-factor between the two spectra is ${\cal R}_{\epsilon M} = 29\%$.  
  }
\label{fig:SK229.1b1.cor.ext.3}
\end{figure}

\begin{figure}
\begin{center}
\resizebox{!}{5cm}{\includegraphics{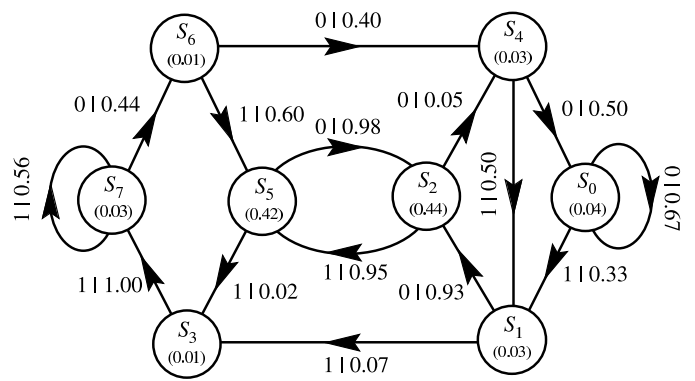}}
\end{center}
\caption{
  The $\Range =3$ \eM\ reconstructed for SK229. The large asymptotic state
  probabilities for the $\CausalState_2$ and $\CausalState_5$ states, as well
  as the large state-transition probabilities between them, show that this is
  predominately a 2H crystal with some faulting.
  }
\label{fig:dB3.229}
\end{figure}

Lastly, we examine an as-grown 2H crystal. The diffraction spectrum for this
crystal is shown in Fig.~\ref{fig:SK229.1b1.cor.ext.3}. We find the
figures-of-merit closest to their theoretical values over the interval
$l \in [-0.33,0.67]$ with values of $\gamma = -0.49$ and  $\beta = 1.00$. 
We find that the smallest-$r$
\eM\ that gives reasonable agreement between the measured and \eM\ spectra
has a memory length of $\Range =3$. The reconstructed \eM\ is shown in
Fig.~\ref{fig:dB3.229}. The large asymptotic state probabilities for the
$\CausalState_2$ and $\CausalState_5$ CSs, as well as the
large inter-state transition probabilities between them, indicate this is
predominantly a 2H crystal. More specifically, the probability of seeing
sequences 1010 and 0101, corresponding to the 2H cycle, have a combined total
weight of about $82.5\%$. The remaining probability is distributed among the
other fourteen length-$4$ sequences. It is tempting to interpret the remaining
structure as faults and, indeed, we can.

Let us treat the transitions $s = 0$ from $\CausalState_6$ and $s = 1$ from
$\CausalState_1$ as though they are missing. These are the least probable
transitions in the \eM:
$\Prob(0,\CausalState_6) = \Prob(\CausalState_6)\Prob(0|,\CausalState_6) \approx 0.004$ and
$\Prob(1,\CausalState_1) = \Prob(\CausalState_1)\Prob(1|,\CausalState_1) \approx 0.002$.
Then, in the left half of the \eM\ there is structure associated with a 2H
deformation fault $[\CausalState_7\CausalState_6\CausalState_5\CausalState_3]$
with probability weight
$0.040 = \Prob(1011) + \Prob(0111) + \Prob(1110) + \Prob(1101)$.
In the right half there likewise is 2H deformation fault structure
$[\CausalState_0\CausalState_1\CausalState_2\CausalState_4]$
with weight
$0.049 = \frac{1}{2}\Prob(0100) + \Prob(1000) + \Prob(0001) + \frac{1}{2}\Prob(0010)$.
The right portion contains the CSC
$[\CausalState_2\CausalState_4\CausalState_1]$ with weight
$0.036 = \frac{1}{2}\Prob(0100) + \Prob(1001) + \frac{1}{2}\Prob(0010)$,
which is associated with growth faults. The CSCs $[\CausalState_0]$ and
$[\CausalState_7]$, identified as 3C structure, have a combined weight of
$0.041$. Given these observations, the interpretation suggested by the \eM\ is
that SK229 has crystal structures and faults in the proportions given in
Table \ref{Table:SK229}. The decomposition there is reasonable since
there is an underlying crystal structure present and the smaller, faulting
paths are not too large. 

\begin{table}
\begin{center}
\begin{tabular}{lr}
\hline
\hline
 Structure         & Contribution \\
 \hline
 2H                &    $82\%$    \\
 3C                &    $ 4\%$    \\
 Deformation fault &    $ 9\%$    \\
 Growth fault      &    $ 4\%$    \\
 Other disorder    &    $ 1\%$    \\
\hline
\hline
\end{tabular}
\end{center}
\caption{The Fault Model structural interpretation of the reconstructed
  \eM\ of Fig.~\ref{fig:dB3.229}. This is valid only under the assumption
  of weak faulting.
  }
\label{Table:SK229}
\end{table}

SK229 has not been analyzed quantitatively using the FM. By comparing the
FWHM of the integer-$l$ to half-integer-$l$ peaks, Sebastian and
Krishna~\cite{sk94} concluded that deformation faulting is the primary
vehicle responsible for the deviation from crystallinity seen here. We are
in agreement, except that we also detect small amounts of 3C crystal structure
and some growth faults. 

\begin{figure}
\begin{center}
\resizebox{!}{5cm}{\includegraphics{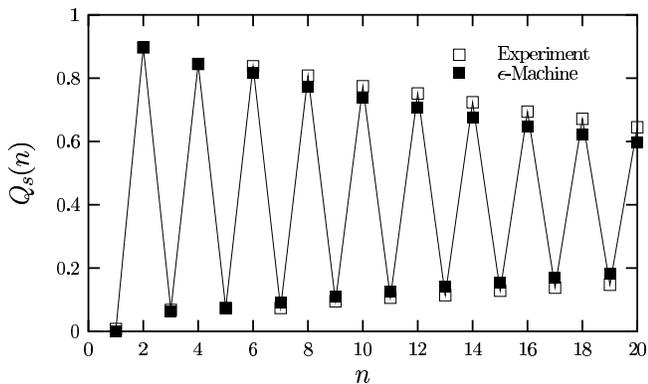}}
\end{center}
\caption{Comparison of the $Q_s(n)$ versus $n$ for experimental spectrum
  SK229 (solid squares) and the \eM\ (open squares).
  }
\label{fig:qsSK229.r.3a}
\end{figure}

Figure~\ref{fig:qsSK229.r.3a} compares the CFs from experiment and the \eM.
The agreement is good, although the reconstructed \eM\ underestimates somewhat
the magnitude of the oscillations in $Q_s(n)$. 
A visual comparison of the experimental diffraction spectrum and that 
generated from the \eM\ (Fig.~\ref{fig:SK229.1b1.cor.ext.3}) shows that there
is reasonable agreement. We calculate an ${\cal R}$-factor between the two
spectra of ${\cal R}_{\epsilon M} = 29\%$.

There are some noticeable differences between the two spectra, however. First,
we see that Bragg-like peaks in the \eM\ spectrum are slightly shifted from
those in the experimental spectrum. Second, the \eM\ spectrum overestimates
the maximum intensity in each peak. Third, the peak profiles are qualitatively
different. For the \eM, the shoulders of the peaks are broader than those of
the experiment and the crowns are narrower. Indeed, the peaks in the 
experimental spectrum appear plateau-like. The sides are very steep and the
crown is rounded. It is not known if this results from instrument resolution
or it is an observable effect. 

This example 
shows that slightly faulted crystals with sharp Bragg-like structures 
can be difficult to analyze. Certainly, the basic crystal structure 
is clear---in this case 2H, but a very fine experimental $l$-mesh is
needed to map each Bragg peak. In contrast, highly disordered spectra with
significant diffuse scattering are less sensitive to experimental details,
such as instrument resolution. For highly diffuse spectra, where the
assumptions underlying the FM break down, \eMSR both in practice (extraction
of correlation information) and in principle (the \eM\ describes {\em any}
amount of disorder) reaches its full potential.   

From the \eM\ we calculate the three length parameters to be
$\Range =3$ ML, $\mathcal{P} = 3.5$ ML, and $\lambda_c = 19 \pm 2$ ML.
The three measures of intrinsic computation are found to be
$\hmu = 0.30$ bits/ML, $\Cmu = 1.8$ bits, and $\EE =  0.89$ bits.   

\section{Physics from \EMs}
\label{EMPhysics}

Now that we have models (\eMs) for the stochastic processes underlying
the observed ML stacking for each sample, we can calculate a range of
structural, computational, and physical characteristics that describe the
stacking patterns and disorder. In Table~\ref{Table:CompLengths} we list
the measures of intrinsic computation and characteristic lengths calculated
for each sample, as well as for three crystal structures for comparison.

\subsection{Characteristic Lengths in Polytypes}

We first note that it was necessary to perform \eMSR\ up to
$\Range=3$ for each spectrum. This is not surprising since the mechanism of
deformation faulting is expected to be important in ZnS, and the minimum
\eM\ on which deformation faulting structure can be modeled is $\Range=3$.  
This implies, of course, a longer memory length that either pure 3C or 2H
structure alone requires. The generalized periods for the 2H and 3C structures
at $\mathcal{P} = 2$ ML and $\mathcal{P} = 1$ ML, respectively, are also much
shorter than for the disordered structures which average $\mathcal{P} = 5.7$
ML. This shows that there is spatial organization over a modest range---$6$
MLs---for {\em disordered} ZnS crystals. This pales in comparison to
crystalline polytypes with repeat distances over 100 MLs, but is still much
larger than the calculated range of inter-ML interactions of $\sim$ 1 ML.   
We note that many of these long-period crystalline polytypes are believed
to be associated with giant screw dislocations that are expressly absent here.
For both $\Range$ and $\mathcal{P}$, the disordered structures have values
much closer to that expected for the 6H structure. In contrast to a perfect
crystal, the correlation lengths are finite rather than infinite.
Interestingly, the sample that has the most stacking disorder (as measured
by $\hmu$), SK137, also has a  comparatively long correlation length. While
this was previously classified as an as-grown 3C crystal by Sebastian and
Krishna,~\cite{sk94} we find that it also contains a significant amount of
stacking sequence associated with the 2H structure. Since we cannot assume
that any of these structures are in equilibrium or the ground state,
we can draw no conclusions about the range of inter-ML interactions.

\subsection{\label{sec:IntrinsicComp}Intrinsic Computation}

Each of the diffraction spectra we analyze also show much stacking disorder.
Even a spectrum that is quite crystalline, like SK134, has a stacking entropy
of $\hmu = 0.50$ bits/ML. Of course, for a crystal the entropy rate is zero
and for the case of completely disordered stacking one would have $\hmu = 1$
bits/ML. If we compare SK134 and SK135, each beginning as a 2H crystal but
annealed at different temperatures, we see that the latter is slightly 
more disordered, as we expect. The statistical complexity, a measure of the 
average history in MLs needed to predict the next ML, is also relatively
constant at $2.3$ bits and $2.5$ bits, respectively.

In fact, the measures of
intrinsic computation are nearly equal, except those for SK229. But SK134 and
SK135 have very different structures. SK134 is largely 2H in character while
SK135 is largely twinned 3C. Assuming that they were identical before annealing,
this would suggest that the disordering process has little effect on these
measures. This is, of course, tentative, since such a conclusion can only
be drawn after examining many disordered samples. Experimental spectra in the
midst of the 2H-to-3C transformation would be of significant interest here.
It is possible, though, that SK137 might be such a instance. While this is an
as-grown twinned 3C crystal, as noted above this, this crystal also has
some significant 2H character. Since Sebastian and Krishna~\cite{sk94} found
that both of these were well described by a random distribution of twin faults,
they concluded that disordered 3C crystals found in the growth furnace result
from a phase transformation from the 2H structure upon cooling the furnace.
We find that the two samples (SK135 and SK137), while similar, do have
qualitative differences. We can understand this either as a crystal not
completely transformed or that the mechanism which created SK137 is not
simple. We feel that more experimental data is needed in order to arrive at
a more complete understanding. Since $\hmu = 0.65$ bits/ML for SK137 and is
thus more disordered than either SK134 or SK135, the interpretation of this
crystal being in the midst of a phase transition is a plausible explanation.
The most striking feature of the measures of intrinsic computation is their
relative consistency (except for SK229) even while the structure of the
crystal changes significantly. 

\begin{table}
\begin{center}
\begin{tabular}{l|lllllll}
\hline
\hline
 System &  $\lambda_c$ & $\mathcal{P}$ & $\Range_l$ & $\hmu$
	& $\Cmu$ &  $\EE$ & $\Delta$ \\
\hline
 2H     &  $\infty$      & 2   & 1  & 0     & 1.0 &  1.0  & 0.0 \\
 3C     &  $\infty$      & 1   & 0  & 0     & 0.0 &  0.0  & 0.0 \\
 6H     &  $\infty$      & 6   & 3  & 0     & 2.6 &  2.6  & 0.0 \\
 SK134  &  $9.5 \pm 0.5$ & 4.8 & 3  & 0.50  & 2.3 &  0.75 & -0.1 \\
 SK135  &  $4.4 \pm 0.7$ & 5.6 & 3  & 0.59  & 2.5 &  0.71 & 0.0 \\
 SK137  &  $12 \pm 3$    & 6.7 & 3  & 0.65  & 2.7 &  0.79 & 0.0 \\
 SK229  &  $19 \pm 2$    & 3.5 & 3  & 0.30  & 1.8 &  0.89 & 0.0 \\
\hline
\hline
\end{tabular}
\end{center}
\caption{A comparison of the three characteristic lengths and three measures
  of intrinsic computation that one can calculate from knowledge of the \eM.
  We calculate them for the experimental diffraction spectra, as well as for
  three crystal structures for reference. Recall that
  $\Delta = \Cmu - \EE - r \hmu$.\cite{vcc02b}
  }
\label{Table:CompLengths}
\end{table}

\subsection{Stacking-Fault Energies}

One physical quantity amenable to calculation from the \eM\ is the difference
in configurational energies of the particular polytypes.
Numerical calculations find that the configurational energy depends only
the nearest and the next-nearest neighbors in the stacking arrangement.
Engel and Needs~\cite{en90} have done a first-principles pseudopotential
calculation of the total energy of five ZnS polytypes, from which
they determined the strength of the interactions up to the third
nearest layer. The most general expression possible for inter-ML
interactions up the third nearest neighbors is given by~\cite{sh90}
\begin{eqnarray}
E & = & E_0 - J_1 \sum_i s_is_{i+1} - J_2 \sum_i s_is_{i+2} \nonumber \\
  &   & - J_3 \sum_i s_is_{i+3} -  K \sum_i s_is_{i+1}s_{i+2}s_{i+3} ~.
\label{eq:stacking_energy}
\end{eqnarray}
Terms with an odd number of spins do not appear due to symmetry considerations.
We take $s_i = \pm 1$ here.

Engel and Needs~\cite{en90} found that $J_1 = 0.00187$ eV per ZnS pair and
$J_2 = -0.00008$ eV per ZnS pair and that $J_3$ and $K$ are negligible.
Given this let us rewrite Eq.~(\ref{eq:stacking_energy}) in terms of the
energy per ZnS pair and take $E_0 = 0$. Then the configurational energy is
\begin{eqnarray}
\tilde{E} = - J_1 \langle s_is_{i+1} \rangle - J_2 \langle s_is_{i+2} \rangle ~,
\label{eq:stacking_energy1}
\end{eqnarray}
where brackets indicate the expectation value over the stacking sequence.
The expectation values are found directly from sequence probabilities,
as follows:
\begin{eqnarray}
\langle s_is_{i+1} \rangle & = & \Prob(11) + \Prob(00) - 2\Prob(01) \nonumber \\
\langle s_is_{i+2} \rangle & = & \Prob(111) + \Prob(101) + \Prob(000) + \Prob(010) \nonumber \\
                           &   & - 2 \Prob(110) - 2 \Prob(100).
\label{eq:stacking_energy_word_probs}
\end{eqnarray}

The configurational energy in terms of meV per Zn-S pair is shown in
Table~\ref{tab:energy_conf} for several crystalline structures and each of
the four disordered samples. The configurational stacking energies are
bounded above and below by the 2H and 3C structures with relative
configurational energies of 1.95 meV/ZnS-pair and -1.79 meV/ZnS-pair,
respectively. For SK134, the annealing process has introduced faults and has
lowered the average stacking energy from the original 2H structure to $1.13$
meV/ZnS-pair, while increasing the stacking entropy. If we assume that SK137
is a partially transformed 2H-to-3C crystal (though mostly 3C), then we see
that the crystal experiences further disordering and the stacking energy
falls to -0.57 meV/ZnS-pair. SK135 shows the most advanced transformation
with the 2H structure almost completely eliminated and stacking energy not
too far from the ideal minimum at -1.02 meV/ZnS-pair. The stacking entropy
begins to fall, however, as the transformation nears a disordered 3C crystal.
Being only a slightly disordered 2H crystal, SK229 shows the highest stacking
energy of 1.56 meV/ZnS-pair. As we might expect from the relative magnitudes of
$J_1$ and $J_2$, the contribution from the $J_1$ term completely dominates
the energy.

\begin{table}
\begin{center}
\begin{tabular}{l|rrrrl}
\hline
\hline
System & $\langle s_is_{i+1} \rangle$ & $\langle s_is_{i+2} \rangle$
  & $\tilde{E}$ & $\alpha_h$ & History \\
\hline
 2H    & -1.00 &  1.00 &  1.95 & 1.00 & PC                  \\
 3C    &  1.00 &  1.00 & -1.79 & 0.00 & PC                  \\
 6H    &  0.33 & -0.33 & -0.65 & 0.33 & PC                  \\
 SK134 & -0.58 &  0.63 &  1.13 & 0.80 & D 2H, 300 C for 1h  \\
 SK135 &  0.56 &  0.45 & -1.02 & 0.24 & D 2H, 500 C for 1h  \\
 SK137 &  0.32 &  0.45 & -0.57 & 0.34 & AG 3C               \\
 SK229 &  -0.80 &  0.86 & 1.56 & 0.90 & AG 2H               \\ 
\hline
\hline
\end{tabular}
\end{center}
\caption{Relative configurational energies $\tilde{E}$ of experimental
  polytypes and
  several pure-crystalline polytypes. The last column gives the history of
  each sample, where PC stands for perfect crystal, AG as-grown, and D
  disordered. 
  We use the energy coupling constants, $J_1$ and $J_2$, calculated by Engels
  and Needs along with the reconstructed \eM\ for the disordered processes
  to find the configurational energy of the disordered structures via
  Eqs.~(\ref{eq:stacking_energy1}) and (\ref{eq:stacking_energy_word_probs}).
  }
\label{tab:energy_conf}
\end{table}

\subsection{Hexagonality}

The degree of birefringence of ZnS crystals is known to depend only
a single structural parameter---the hexagonality $\alpha_{h}$.~\cite{bs66}
This parameter is defined as that fraction of MLs which are hexagonally
related to their neighbors. That is, $\alpha_{h}$ is defined as the frequency
of occurrence of sequences $ABA$ and $BAB$ and their cyclic permutations. In
terms of the H\"{a}gg notation, these are simply $\Prob(01)$ and $\Prob(10)$,
respectively. Since $\Prob(01) = \Prob(10)$,~\cite{vcc02b} we have
\begin{eqnarray}
   \alpha_{h} = 2 \Prob(01) ~.
\label{eq:hexagonality}
\end{eqnarray}
Sequence probabilities are directly calculable from the \eM, so that the
hexagonality of a disordered crystal can be easily found.

In Table~\ref{tab:energy_conf} we show the hexagonality calculated for all
of the spectra, as well as for several crystal structures for comparison.

\section{Discussion}
\label{Discussion}

We have successfully applied computational mechanics to the discovery
and description of stacking order in single crystals of polytypic ZnS.
In doing so, we reconstructed from experimental diffraction spectra the
minimal, optimal, and unique description of the stacking process as
embodied in the \eM. In contrast to previous analyses,~\cite{sk94} we
used \emph{all} of the information in the diffraction spectra, both in
the Bragg peaks and in the diffuse scattering between them. We imposed
no restrictions on the kind of structures to be found, save that they be
representable by \eMs. Further, the computational mechanics approach was
not limited to the case of weak faulting, but can be used to treat even
highly disordered samples. Additionally, the \eM\ can naturally accommodate
more than one parent crystalline structure as seen in SK134.

For two of the spectra, a sensible decomposition~\cite{note3} 
of the \eM\ into crystal and faulting structure
was possible, allowing a direct comparison between the computational
mechanics approach and the FM. The \eM\ detected structures not previously
found by the FM. For example, in SK134
we found that not only was structure associated with deformation
faulting important, but there was also structure related to growth
faults and layer-displacement faulting. We even found nascent
sequences leading to the 3C structure. For the other two cases, while no
FM-like decomposition of the \eM\ was proposed, we still found significant 
structure as embodied in the \eM.  
From the \eM, physical 
insight into the structure of the stacking was possible. For example, in the
$\Range =3$ reconstructed \eM\ for SK137 we could eliminate
6H structure based on the absence of a transition between CSs. We
also found that 2H structure was present. Even when no sensible
decomposition into a simple pure-crystal and weak-faulting structure
is possible, the \eM\ still directly provides sequence frequencies,
which can be used to build physical insight into the stacking structure.

From a knowledge of the \eM, it is possible to calculate a number of
physical characteristics. In Table~\ref{tab:energy_conf} we
tabulated the stacking entropy per ML for each spectrum.
Given the coupling parameters between MLs we calculated
the average stacking-fault energy for the samples, as shown
in Table~\ref{tab:energy_conf}. We were also able to find the
hexagonality for the disordered crystals. Knowing the \eM\ allowed
us to find various characteristic lengths associated with each
disordered crystal, such as the memory length and the generalized
period.  We believe that additional physics can be calculated
from the \eM.

We also calculated measures of intrinsic computation from the 
\eM. We found that the minimum memory length for all samples
was $\Range =3$, which is in excess of the calculated inter-ML interaction
range of $\sim 1$ ML for ZnS. We further found that the statistical complexity,
a measure of the amount of information in the \eM, was also much larger than
that of either the pure 2H or 3C structures. Also, the range over which
structures are found in the disordered samples is about 6 ML. 

Characterizing solid-state transitions in polytypic materials is of
considerable interest. Let us review what an \eM\ does and does not imply.
Most simply put, the \eM\ is the answer to the question, What is the minimal,
optimal, and unique description of the one-dimensional stacking structure
of the sample?Any physical parameters that depend on this description are
\emph{in principle} calculable from \eM. The \eM\ does \emph{not} answer
the question, How did the crystal come to be stacked in this way? To
determine this, one must augment structural knowledge (as embodied in the \eM)
with additional information or assumptions. Such information can come in the
form of a time series of structures obtained either from a series of numerical
simulations or experiments or, in the theoretical domain, perhaps from
assumptions about weak faulting.

Since we have discovered structures in polytypic ZnS that were undetected
before, we feel that the mechanism of faulting---previously attributed
to deformation faulting---deserves re-examination. We have provided
a firm theoretical foundation for the discovery and description of
disordered stacking sequences in polytypes and believe that additional
experimental studies are warranted. Additionally, computer simulations
of solid-state transformations in polytypes with proposed faulting
mechanisms, accompanied by the concomitant reconstruction of the
\eM\ directly from the sequence of MLs from the
simulation, should provide a powerful method to understand the
gross features of the transformations.

Before such studies can be definitive, a quantitative understanding
of the effects of experimental error on the reconstructed \eM\ is needed.
With the exception of our introduction of the figures-of-merit, $\gamma$
and $\beta$, we have not addressed this important issue here. We note that
the original experimental data was not reported with error bars and this means
that comparison with the desired \eM\ error analysis would have not been
possible. Additionally, the necessity to digitize the data undoubtedly
introduced errors. It is therefore difficult to assess the amount of error in
each spectrum.  Our intuition tells us that error in the diffraction spectrum
will likely lead to suppression of the more delicate structures on an \eM.
Therefore, one should expect to find \emph{less} structure and \emph{more}
randomness.

We mention that the application of computational mechanics to the
description of one-dimensional sequences is the most general approach
possible to this problem. Thus, its application here to polytypism
represents the end point in the evolution of models to describe
the disordered sequences seen these substances. Any alternate description
can be expressed as an equivalent \eM\ and none can be more general,
since in the language of statistics an \eM\ is the minimal sufficient
statistic for the underlying process. It may be possible to find specialized
algorithms that are more sensitive or more efficient in determining an
\eM\ than the one we introduced in \eM\ spectral reconstruction. However,
the answer, in its most general form, will be expressible as an \eM.

We also note that our work here represents a solution to a significant
theoretical problem---How does one extract structural information
from a power spectrum? Our application has been to polytypism, but
the principles underlying our solution may be applied quite generally
to domains in which spectral information is available.

Future directions for this work include an application of \eMSR\ to other
polytypes, as well as to substantially more complex materials. The extension
of these ideas to the more common cases of disorder in two and three
dimensions is also desirable. The development of computational mechanics in
higher dimensions would significantly aid in the classification and
understanding of disorder in many physical systems.
 
\begin{acknowledgments}

This work was supported at the Santa Fe Institute under the Networks
Dynamics Program funded by the Intel Corporation and under the
Computation, Dynamics, and Inference Program via SFI's core grants
from the National Science and MacArthur Foundations. Direct support
was provided by NSF grants DMR-9820816 and PHY-9910217 and DARPA
Agreement F30602-00-2-0583.  DPV's original visit to SFI was supported
by the NSF.

\end{acknowledgments}


\bibliography{master.refs}

\begin{thebibliography}{57}
\bibitem[{not({\natexlab{a}})}]{note2}
\bibinfo{note}{We use the Ramsdell notation to specify common crystal
  structures in CPSs; see Ref. \onlinecite{vcc02b}}.
\bibitem[{not({\natexlab{b}})}]{note4}
\bibinfo{note}{Notation and definitions of variables used here are introduced
  elsewhere; see Ref. \onlinecite{vcc02b}}.
\bibitem[{not({\natexlab{c}})}]{note3}
\bibinfo{note}{Recall from Part I\cite{vcc02b} that this decomposition is not
  unique due to assumptions used in the FM.}
\end{thebibliography}

\end{document}